\documentclass[fdp,a4paper,fleqn]{w-art}
\usepackage{times,cite,w-thm}
\theoremstyle{plain}

\theoremstyle{definition}

\usepackage[]{graphicx}
\begin{document}
\DOIsuffix{theDOIsuffix}
\Volume{55}
\Month{01}
\Year{2007}
\pagespan{1}{}
\Receiveddate{XXXX}
\Reviseddate{XXXX}
\Accepteddate{XXXX}
\Dateposted{XXXX}
\keywords{Porter-Thomas distribution, random matrix theory.}


\title[Neutron Data Exclude RMT]{Neutron Resonance Data Exclude 
Random Matrix Theory}

\author[P.E. Koehler]{P.E. Koehler\inst{1,}%
  \footnote{Corresponding author\quad E-mail:~\textsf{koehlerpe@ornl.gov},
            Phone: 01\,865\,574\,6133,
            Fax: +01\,865\,576\,8746}}
\address[\inst{1}]{Physics Division, Oak Ridge National Laboratory, MailStop
6356, Oak Ridge, Tennessee 37831, USA}

\author[F.~Be\v{c}v\'{a}\v{r}]{F.~Be\v{c}v\'{a}\v{r}\inst{2}}
\address[\inst{2}]{Charles University, Faculty of Mathematics and Physics,
180 00 Prague 8, Czech Republic}

\author[M. Krti\v{c}ka]{M. Krti\v{c}ka\inst{2}}

\author[K. H. Guber]{K. H. Guber\inst{3}}
\address[\inst{3}]{Reactor and Nuclear Systems Division, Oak Ridge National
Laboratory,Mail Stop 6356, Oak Ridge, Tennessee 37831, USA}

\author[J. L. Ullmann]{J. L. Ullmann\inst{4}}
\address[\inst{4}]{Los Alamos National Laboratory, Los Alamos, New Mexico
87545, USA}

\begin{abstract}
Almost since the time it was formulated, the overwhelming consensus 
has been that random matrix theory (RMT) is in excellent agreement with 
neutron resonance data. However, over the past few years, we have 
obtained new neutron-width data at Oak Ridge
and Los Alamos National Laboratories that are in stark disagreement with
this theory. We also have reanalyzed neutron widths in the most famous data
set, the nuclear data ensemble (NDE), and found that it is seriously 
flawed, and, when analyzed carefully,
excludes RMT with high confidence. More recently, we carefully
examined energy spacings for these same resonances in the NDE using the $\Delta _{3}$
statistic. We conclude that the data can be found to either confirm or refute
the theory depending on which nuclides and whether known or suspected 
$p$-wave resonances are included in the analysis, in essence confirming
results of our neutron-width analysis of the NDE. We also have
examined radiation widths resulting from our Oak Ridge and Los Alamos
measurements, and find that in some cases they do not agree with RMT. 
Although these disagreements
presently are not understood, they could have broad impact on basic and
applied nuclear physics, from nuclear astrophysics to nuclear criticality
safety.
\end{abstract}

\maketitle

\section{Introduction}

In 1956, Porter and Thomas \cite{Po56} proposed a theory to explain the
surprising discovery that $s$-wave resonance reduced neutron widths 
($\Gamma _{n}^{0}$) 
spanned a very wide range. Starting from seemingly sound and
fundamental assumptions, their theory predicted that $\Gamma _{n}^{0}$
values follow a $\chi ^{2}$ distribution with one degree of freedom 
($\nu_{n}=1$), which subsequently became known as the Porter-Thomas distribution
(PTD).

Almost since that time, the overwhelming consensus has been that data and
theory agree very well. In fact, faith in this theory is so strong that in
the past $\sim $30 years it has been extremely rare to find a paper in
the literature in which new data have been used to test the theory. Instead,
standard procedure has been to use the theory to correct new data for
experimental deficiencies. In the intervening years, random matrix theory
(RMT) \cite{We2009} was developed and has placed the PTD on more formal
footing, broadened the scope and predictions, and provided links between
nuclear physics and many other fields, including quantum chaos. As a
consequence, the impact of neutron resonance data has become much broader,
as such data are routinely cited as some of the best proof of the veracity
of RMT.

In this paper, we summarize recent tests of RMT predictions using 
neutron resonance data. Because all experiments from which such data 
are obtained have important limitations, we begin by discussing these 
problems and how they are circumvented in Section \ref{ExpLimits}.

Over the past few years, we have obtained new $\Gamma _{n}^{0}$ data 
\cite{Ko2007,Ko2010a} at the Oak Ridge Electron Linear Accelerator (ORELA) 
\cite{Pe82} and Los Alamos Neutron Science Center (LANSCE) \cite{Li90} 
facilities that are much better than previous data. We discuss results 
of using these data to test the PTD in Section \ref{PTDTests}.

The $\mathcal{R}$-matrix analyses from which $\Gamma _{n}^{0}$ 
data are obtained also yield total radiation widths ($\Gamma _{\gamma }$) for
these same resonances. These data also can be used to test the PTD and other
predictions of theory, but the situation is complicated by 
limited knowledge of the level density as a function of excitation 
energy and the photon strength functions needed to model the $\gamma$ decay  
of the resonant capturing states. In Section \ref{GgTests}, we discuss some
general observations regarding what has been learned from $\Gamma _{\gamma }$
data obtained at ORELA over the past several years, as well as a surprising
very recent result \cite{Ko2012} discerned from our $^{147}$Sm data taken at 
LANSCE.

In addition to our own data, we have reanalyzed \cite{Ko2011} neutron widths
in the most famous data set \cite{Ha82,Bo83}, the so-called nuclear data
ensemble (NDE). More recently, we have reanalyzed energy spacings for these
same resonances, with careful attention to evaluation of uncertainties. We
discuss results of these analyses in Section \ref{NDE}.

In almost all cases we have examined so far, we find significant deviations
from the PTD or other disagreements with RMT predictions. There also
have been several other reported disagreements with the PTD \cite%
{Ri69,Fo71,Fo71a,Ra72,Ca76}, which we briefly describe in Section 
\ref{OtherExps}. To our knowledge, those results have never been
explained or retracted.

As far as we know, these disagreements with RMT presently are
not understood, although our recent results have (re)ignited theoretical
interest \cite{We2010,Ce2011,Vo2011,Sh2012}, some of which is closely
related to earlier work \cite{Kl85} on non-Hermitian Hamiltonians. 
We very briefly discuss recent theoretical work in Section \ref{Theory}. Because
many models assume that both particle and primary gamma transitions follow
the PTD, violation of this assumption could have broad impact on basic and
applied nuclear physics, from nuclear astrophysics to nuclear criticality
safety.

Reliable separation of small $s$- from large $p$-wave resonances remains 
the largest barrier to better tests of RMT using neutron resonance 
data. In Section \ref{NewExpTechs}, we very briefly describe new experiment
techniques aimed at surmounting this barrier.

We end with a short summary and some conclusions in Section \ref{Conclusions}.

\section{Experiment limitations}\label{ExpLimits}

\begin{figure}[b]
\includegraphics[clip,width=0.6\columnwidth]{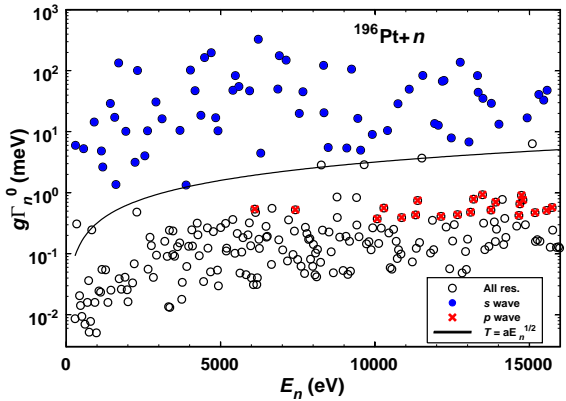} 
\vspace*{-0.3cm}
\caption{Effective reduced neutron widths versus energy 
for $^{196}$Pt resonances from our ORELA data. Open circles, 
filled circles, and X's depict data for all, 
firm $s$-, and firm $p$-wave resonances, respectively. The 
threshold used in the ML analysis of Ref. \cite{Ko2010a} is 
shown as a solid curve.}
\label{Pt196gGn0VsE}
\end{figure}

An ideal test of the PTD or other predictions of RMT would involve a
complete (no missing resonances) and pure (all resonances having the same
parity) data set. Typically, $s$-wave neutron resonances are used
because they are largest and hence easiest to observe. Unfortunately, all
experiments have a lower limit for observing resonances as well as a
threshold below which $s$- and $p$-wave resonances cannot be
differentiated. These facts are illustrated in Fig. \ref{Pt196gGn0VsE}, in 
which neutron widths for $^{196}$Pt resonances from our ORELA measurements 
are plotted as a function of resonance energy.

To understand how the data in this figure 
illustrate these facts, it is useful to know the following details. 
Because $^{196}$Pt is a 
zero-spin target ($I^{\pi}=0^{+}$), all $s$-wave resonance have spin and 
parity $J^{\pi}=\frac{1}{2}^{+}$, and statistical spin factor 
$g_{J}=\frac{(2J+1)}{(2I+1)(2j+1)}=1$, where $j=\frac{1}{2}$ is the neutron
spin. In contrast, there are two spin/parity values, $\frac{1}{2}^{-}$ and 
$\frac{3}{2}^{-}$, for $p$-wave resonances, for which $g_{J}=1$ and 
2, respectively. Shown in Fig. 
\ref{Pt196gGn0VsE} are effective reduced neutron widths 
$g\Gamma_{n}^{0}=g\Gamma_{n}/\sqrt{E_n/\mathrm{1 eV}}$. For $s$-wave resonances, these are 
the usual reduced neutron widths which should, on average, be constant. 
However, for $p$-wave resonances, these effective reduced widths should be, 
on average, proportional to $E_{n}$. Also, according to standard level-density 
models, the average number of resonances should be proportional to $2J+1$ 
and independent of parity. Therefore, it is expected that there should 
three times as many $p$- as $s$-wave resonances (i.e., $D_{0}=3D_{1}$). 
Finally, due to the overall larger penetrability for $s$ waves and the fact 
that Pt is near the peak of the $s$- and valley of the $p$-wave neutron 
strength function, $s$-wave resonances should be much larger than $p$-wave 
ones at these energies.

These expectations are in qualitative agreement with the data in Fig. 
\ref{Pt196gGn0VsE}. For example, it is evident that there is 
one group of resonances with larger and constant (on average) $g\Gamma_{n}^{0}$ 
and another, more populous group, having smaller effective reduced neutron 
widths with sizes (on average) proportional to $E_{n}$. We were able to make 
firm $s$-wave assignments to almost all resonances in the former group by 
virtue of their asymmetric shape (due to 
interference with the comparatively large $s$-wave potential background) in 
the total cross section. Such resonances are depicted by solid symbols in 
this figure. Also, we were 
able to make a few firm $J^{\pi}=\frac{3}{2}$ (and hence $p$-wave) 
assignments under the assumption that the total $\gamma$ width 
$\Gamma_{\gamma}$ does not vary by more than a factor of three. These 
resonances are marked by X's in Fig. \ref{Pt196gGn0VsE}. Unfortunately, 
both techniques for assigning resonance parity cease to work when the 
neutron width becomes too small. Nevertheless, it is obvious from this 
figure that if an analysis threshold shown by the solid curve is used, there 
is only an extremely small chance that a $p$-wave resonance will be included. 
However, it is still likely that some $s$-wave resonances are below this 
threshold, so this fact must be accounted for in any subsequent analysis. 
These data represent some of the best ever acquired for the purpose of 
testing RMT; by virtue of the excellent sensitivity for observing small 
resonances, the good separation of $s$- from $p$-wave resonances, 
and the number of firm parity assignments. Even so, it is vital to 
realize that a pure and complete set of $s$-wave resonances still could not be 
obtained.

To avoid systematic errors, these experiment facts must be 
incorporated into the analysis
technique. These problems affect $\Gamma _{n}^{0}$, $\Gamma _{\gamma }$, and
resonance-spacing data in different ways, so each will be discussed
separately below.

\section{Testing the PTD with $\Gamma _{n}^{0}$ data}\label{PTDTests}

The maximum-likelihood (ML) technique is most often used to test whether
data are consistent with the PTD because it is statistically efficient.
Because the PTD is a special case ($\nu_{n} =1$) of the family of $\chi ^{2}$
distributions, it is assumed that reduced neutron widths are distributed
accordingly and the ML method is used to estimate the most likely value of $%
\nu_{n} $. Given the shape of the $\chi ^{2}$ distribution as a function of $\nu_{n} 
$, neglecting the effect of missed $s$-wave resonances below
threshold will, in general, lead to a falsely large value of $\nu_{n} $ from the
ML analysis. Conversely, including even just a few \textit{p}-wave
resonances in an $s$-wave set will, in general, lead to a falsely
small value of $\nu_{n} $.

Typically, these difficulties have been surmounted by using an
energy-independent threshold as an integral part of the ML analysis,
implicitly assuming that all $s$-wave resonances above threshold have
been observed. We recently have shown \cite{Ko2010a} that an
energy-dependent threshold (on $\Gamma _{n}^{0}$) of the form $T=aE_{n}$,
where $a$ is a constant factor and $E_{n}$ the neutron energy, offers three
advantages compared to using an energy-independent threshold. First, \textit{%
p}-wave contamination is eliminated equally effectively at all energies.
This is because the penetrability factor for \textit{p} waves differs from 
$s$ waves by (to good approximation) a factor of $E_{n}$. Second,
experiment thresholds have approximately this same energy dependence; thus
possible diffusiveness of the instrumental threshold can be surmounted
equally effectively at all energies. Third, statistical precision of the
analysis is maximized by allowing the largest \textit{p}-wave-free set of 
$s$-wave resonances to be included.

The analysis technique was described in Refs. \cite{Ko2010a,Ko2011}. Each
resonance $\lambda $ has an energy $E_{\lambda }$ and reduced neutron width $%
\Gamma _{\lambda n}^{0}$. The probability density function (PDF) $f(x|\nu_{n} )$
for a $\chi ^{2}$ distribution is given by:
\begin{equation}
f(x|\nu_{n} )dx=\frac{\nu_{n} }{2G(\frac{\nu_{n} }{2})}\left( \frac{\nu_{n} x}{2}\right) ^{%
\frac{\nu_{n} }{2}-1}\exp \left( -\frac{\nu_{n} x}{2}\right) dx,
\end{equation}%
where $G(\frac{\nu_{n} }{2})$ is the gamma function for $\frac{\nu_{n} }{2}$,  
$x\rightarrow \Gamma _{\lambda n}^{0}/\langle \Gamma _{n}^{0}\rangle $, and 
$\langle \Gamma _{n}^{0}\rangle $ is the average reduced neutron width.

The joint PDF for statistical variables $\Gamma _{\lambda \mathrm{n}}^{0}$
and $E_{\lambda }$ is defined in a~2D region defined by
inequalities $E_{\lambda }<E_{\mathrm{max}}$ and $\Gamma _{\lambda \mathrm{n}%
}^{0}>T(E_{\lambda })$, where $E_{\mathrm{max}}$ is an~upper limit of
energies $E_{\lambda }$. The expression for this PDF reads 
\begin{equation}
h^{0}\!\left( E_{\lambda },\Gamma _{\lambda n}^{0}\,|\,\nu_{n} ,\langle \Gamma
_{n}^{0}\rangle \right) =Cf\!\left(  \left. \frac{\Gamma _{\lambda \mathrm{n}}^{0}}{%
\langle \Gamma _{n}^{0}\rangle } \right\vert \nu_{n} \right) .
\end{equation}%
The factor $C$, ensuring a~unit norm of $h^{0}$, is $\nu_{n} $- and $\langle
\Gamma _{n}^{0}\rangle $-dependent. The ML function was calculated from all $%
n_{0}$ pairs $\left[ E_{\lambda _{i}}^{\mathrm{\,exp}},\Gamma _{\lambda _{i}%
\mathrm{n}}^{\mathrm{\,exp}}\right] $ within the specified region 
obtained from the experiment. Specifically, 
\begin{equation}
L\left( \nu_{n} ,\langle \Gamma _{n}^{0}\rangle \right)
=\prod_{i=1}^{n_{0}}h^{0}\!\left( E_{\lambda _{i}}^{\mathrm{\,exp}},\Gamma
_{\lambda _{i}\mathrm{n}}^{\mathrm{0\,exp}}\,|\,\nu_{n} ,\langle \Gamma
_{n}^{0}\rangle \right) .
\end{equation}

One problem with many previous tests of the PTD is that they were made using
data from nuclides near the peaks of the \textit{p}- and valleys of the 
$s$-wave neutron strength functions. As a result, neutron widths for
the two parities are nearly equal and hence it is very difficult, 
if not impossible, to obtain a
pure $s$-wave set spanning a sufficiently wide range of neutron
widths to be useful for testing the PTD. For this reason, it is much better
to work with nuclides near the peaks of the $s$- and valleys of the 
\textit{p}-wave neutron strength functions. We recently have made
measurements on several such nuclides; $^{147}$Sm and $^{192,194,195,196}$Pt.

\begin{figure}[b]
\includegraphics[clip,width=0.6\columnwidth]{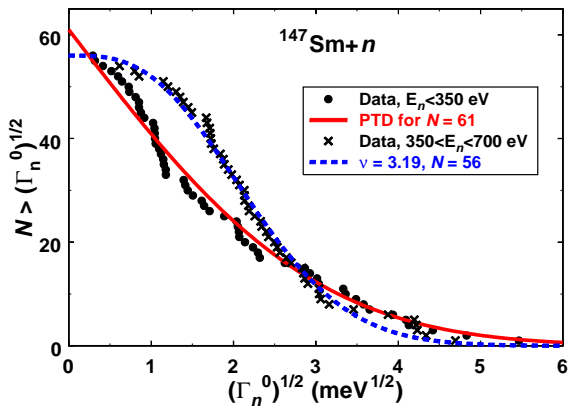} 
\vspace*{-0.3cm}
\caption{Cumulative $\Gamma _{n}^{0}$ distributions from our 
LANSCE data for $^{147}$Sm resonances for two energy regions. 
Shown are the number of resonances having reduced neutron widths 
larger than some value, versus that value. 
Circles and X's depict data for $E_{n}<350$ and 
$350<E_{n}<700$ eV, respectively. Solid and dashed curves depict 
the PTD and a $\chi^{2}$ distribution for $\nu=3.19$, respectively. 
See text and Ref. \cite{Ko2007} for details.}
\label{NW2ERegsBothJs2}
\end{figure}

Our $^{147}$Sm($n,\gamma $) measurements \cite{Ko2007} were made using the
Detector for Advanced Neutron Capture Experiments (DANCE) \cite{Re2004} at
LANSCE. One problem working with odd-A nuclides is that there are two
possible $s$-wave spin states ($J=3$ and 4 in the case of $^{147}$%
Sm), and performing the best test of the PTD requires knowing all the
resonance spins. Using $\gamma $-ray multiplicity data obtained with DANCE,
we developed \cite{Ko2007} and improved \cite{Be2011} an new method for 
separating the two 
$s$-wave spins, and hence determined spins for nearly all observed
resonances for $E_{n}<1$ keV. This allowed us to use the combined $J=3$ and
4 data to test the PTD. As shown in Fig. \ref{NW2ERegsBothJs2}, the 
surprising result was that the data changed from
agreeing very well with the PTD ($\nu_{n} =0.91\pm 0.32$) for resonances below
350 eV, to significant disagreement with the PTD ($\nu_{n} =3.19\pm 0.83$) for
resonances within the next 350 eV. This change occurs at the same energy as
a previously observed \cite{Ko2004} non-statistical behavior in the $\alpha $%
-particle strength function ratio in this nuclide.

Our $^{192,194,195,196}$Pt data \cite{Ko2010a} were taken at ORELA. Neutron
capture and total cross sections for isotopically-enriched samples were
measured using a pair of C$_{6}$D$_{6}$ detectors and a $^{6}$Li-loaded
glass detector, respectively. The neutron total cross section for a $^{nat}$%
Pt sample also was measured. Resonance parameters were determined from a
simultaneous $\mathcal{R}$-matrix analysis of the nine sets of capture and
total cross sections. Parameters for a total of 1264 resonances were
determined, 631 of which could be assigned as $s$ wave, accounting for 
nearly all resonances
above the thresholds used in the ML analyses. The $^{195}$Pt data were 
excluded from the main ML analysis because spins could not be determined 
for enough resonances. The $^{192,194}$Pt data each exclude the 
PTD by nearly three
standard deviations ($\nu_{n} =0.57\pm 0.16$ and $0.47\pm 0.19$, respectively).
Results for $^{196}$Pt were consistent with $^{192,194}$Pt, albeit with
reduced statistical precision ($\nu_{n} =0.60\pm 0.28$) due to the smaller
number of resonances observed for this isotope. Combined in a very
conservative manner, the $^{192,194}$Pt data exclude the PTD with at least
99.997\% confidence.

\section{Testing the PTD with $\Gamma _{\protect\gamma }$ data}\label{GgTests}

The total radiation width $\Gamma _{\gamma }$ for a neutron resonance 
is the sum of partial widths 
$\Gamma _{i\gamma }$ for the different channels by which the capturing state 
can decay by $\gamma$ emission, $\Gamma _{\gamma }=\sum_{i=1}^{n}\Gamma _{i\gamma }$.
According to RMT, partial widths $\Gamma _{i\gamma }$ follow 
the PTD ($\nu _{i\gamma }=1$). In contrast to the neutron case where there
is only a single open neutron channel (elastic scattering) at the relevant energies,
there are many open channels through which $\gamma $ decay can
proceed. Due to the properties of $\chi^{2}$ distributions, in the simplest 
model total radiation widths are predicted to follow a $\chi ^{2}$
distribution with degrees of freedom given by the {\em effective} number of
independently-contribution channels, $n_\mathrm{eff} \equiv \nu _{\gamma
}=\sum_{i=1}^{n}\nu _{i\gamma }$. As $n_\mathrm{eff} \sim 100$, 
$\Gamma _{\gamma }$ distributions are predicted to be very narrow. 
Moving beyond this simplest theory requires detailed simulation of the 
partial radiation widths for primary transitions, which in turn requires 
accurate knowledge of the relevant level densities and photon strength 
functions. On the other hand, because $\Gamma _{\gamma }$ distributions are
much narrower than the PTD, it might be easier to detect deviations from
theory, especially given the limited number of resonances typically
available.

\begin{figure}[b]
\includegraphics[clip,width=1.0\columnwidth]{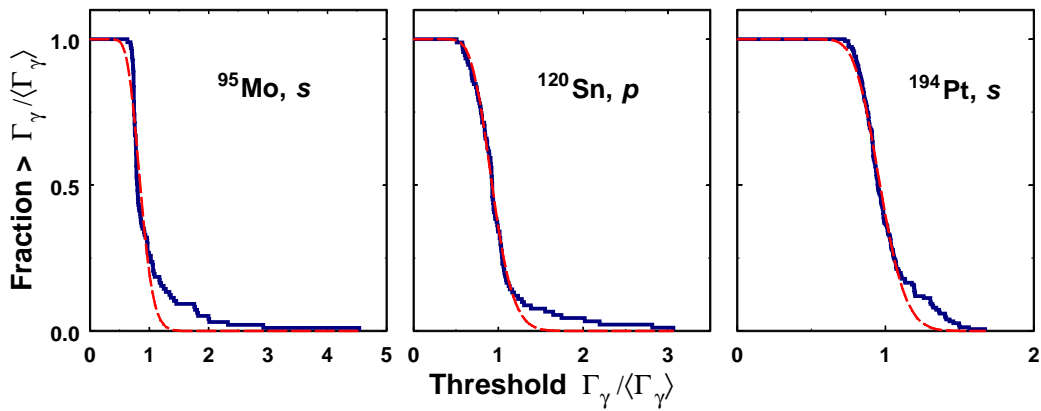} 
\vspace*{-0.3cm}
\caption{Cumulative $\Gamma _{\gamma }$ distributions from our 
ORELA data for three different nuclides. Shown are the fraction of 
resonances with $\Gamma _{\gamma }/\langle \Gamma _{\gamma }\rangle$ 
larger than a given value versus the value, where 
$\langle \Gamma _{\gamma }\rangle$ is the average $\Gamma _{\gamma }$ 
determined from the data. Solid staircase plots depict measured 
data whereas dashed smooth curves show $\chi ^{2}$ distributions.}
\label{ThreeGgDistInLine}
\end{figure}

Comparisons \cite{Ko2011a} we have made to date using our ORELA
data for Pt isotopes indicate that measured $\Gamma _{\gamma }$ 
distributions are broader
than those predicted using standard photon-strength-function and
level-density models. In addition, as shown in in Fig. \ref{ThreeGgDistInLine} 
distributions for several nuclides measured at ORELA appear to
contain extra tails. Although we have observed that 
simulations within the framework of the nuclear statistical model can 
lead to significant differences from $\chi^{2}$ distributions predicted 
by the simplest model, we have been unable to simulate tails as large as 
those shown in Fig. \ref{ThreeGgDistInLine}. Hence, these tails may be the 
sign of non-statistical or collective effects.

\begin{figure}
\begin{minipage}{72mm}
\includegraphics[width=\linewidth]{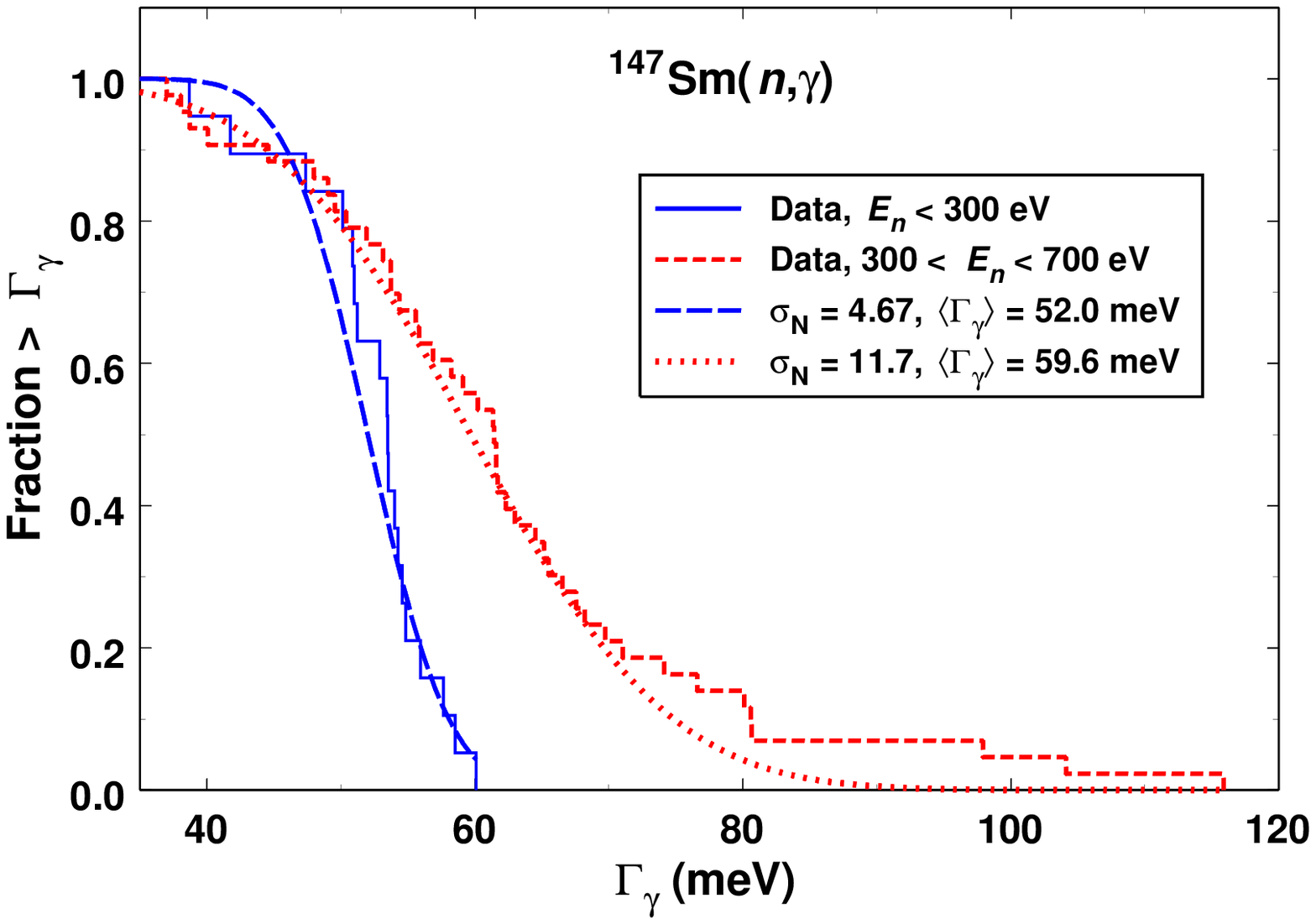}
\caption{Cumulative $\Gamma _{\protect\gamma }$ distributions for neutron 
resonances in $^{147}$Sm. Shown are the
fraction of resonances with $\Gamma _{\protect\gamma }$ larger than a given
value versus the value. Staircase plots depict the measured data whereas
smooth curves show Gaussian distributions from ML analyses. See text and Ref.
\cite{Ko2012} for details.}
\label{GgCumDists2ERegs}
\end{minipage}
\hfil
\begin{minipage}{72mm}
\includegraphics[width=\linewidth]{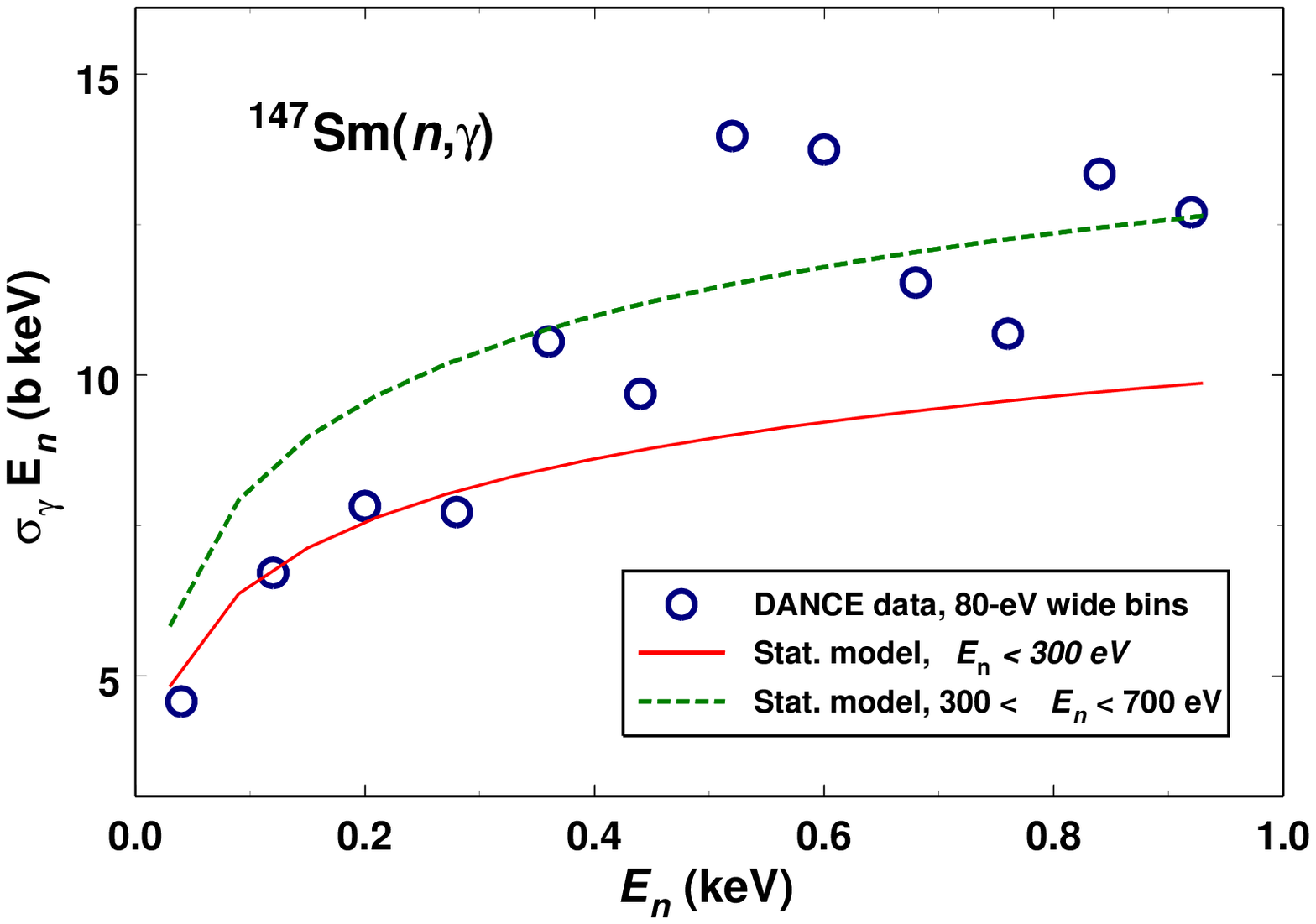}
\caption{Open circles depict our DANCE $^{147}$Sm($n,\protect\gamma $)
cross sections averaged over 80-eV-wide bins. Error bars corresponding to
one-standard-deviation statistical uncertainties are smaller than the
symbols. The solid and dashed curves show results of statistical
model calculations based on the average resonance parameters for the lower-
and upper-energy regions, respectively. See Ref. \cite{Ko2012} for details.}
\label{AveCSvE}
\end{minipage}
\end{figure}

Also, as we very recently reported in Ref. \cite{Ko2012} and illustrated in
Fig. \ref{GgCumDists2ERegs}, both the average total
radiation width $\langle \Gamma_{\gamma} \rangle $ and the width of the $%
\Gamma _{\gamma }$ distribution increase rather abruptly near $E_{n}=300$ eV
for resonances in $^{147}$Sm. 
The median (variance) test \cite{Co80} indicates the null hypothesis that medians
(variances) of the $\Gamma _{\gamma }$ distributions in the two energy
regions are the same can be rejected at the 99.8\% (99.9\%) confidence
level. In addition, the Smirnov and Cramer-von Mises two-sample 
tests \cite{Co80} reveal
the null hypothesis that data in the two energy regions were sampled from
the same distribution can be rejected with $>99\%$ and $>99.9\%$ confidence,
respectively. In essence, all these statistical tests indicate that the
change in the $\Gamma _{\gamma }$ distribution evident in Fig. \ref%
{GgCumDists2ERegs} is highly statistically significant.

Theoretical interpretation of this change may be aided by estimation of
distribution parameters for the two regions. To this end, we used the ML
method. As noted above, in the simplest model $\Gamma _{\gamma }$ data 
are expected to follow a $\chi ^{2}$ distribution with many degrees of 
freedom, typically $\nu _{\gamma }\sim 100 $. For such large values of 
$\nu _{\gamma }$, a $\chi ^{2}$ distribution
is very close to Gaussian in shape. One advantage of using a Gaussian rather
than $\chi ^{2}$ distribution for the analysis is that data uncertainties 
$\Delta\Gamma _{\gamma }$ can easily be included \cite{As66}.

Therefore, we used the ML technique described in Ref. \cite{As66} to estimate
most likely values for the means $\langle \Gamma _{\gamma }\rangle $ and
standard deviations $\sigma _{N}$ of the $\Gamma _{\gamma }$ distributions
in the two energy regions. Resulting ML estimates are $\sigma _{N}=4.67\pm
0.81$, $\langle \Gamma _{\gamma }\rangle =52.0\pm 1.1$, and $\sigma
_{N}=11.7\pm 1.5$, $\langle \Gamma _{\gamma }\rangle =59.6\pm 2.0$, for the
lower- and upper-energy regions, respectively. Hence, these ML results, 
which take into account the measurement uncertainties, also
indicate that $\Gamma _{\gamma }$ distributions in the two energy regions
are significantly different. Translated to $\chi ^{2}$ distributions, these
ML results yield $\nu _{\gamma }=248\pm 87$ and $52\pm 14$ 
for the $\Gamma _{\gamma }$ distributions in the lower and upper energy 
regions, respectively.

These changes in the $^{147}$Sm $\Gamma _{\gamma }$ distribution occur at
the same energy that the neutron-width distribution changes for resonances
in this nuclide, as described in Section \ref{PTDTests} above. Furthermore, 
as shown in Fig. \ref%
{AveCSvE}, these changes in the $\Gamma _{n}^{0}$ and $\Gamma _{\gamma }$
distributions appear to be mirrored by increases in the both the average $%
^{147}$Sm($n,\gamma $) cross section and fluctuations around the average
near this same energy. Explaining these energy-dependent effects represents
a significant challenge to theory.

\section{Testing RMT with the NDE}\label{NDE}

The NDE \cite{Ha82,Bo83} is a set of 157 proton and 1250 neutron resonance 
energies consisting of 30 sequences in 27 different medium-weight and heavy
nuclides. The ensemble was assembled to test
predictions of RMT. Fluctuation properties of resonance energies in the NDE
were found to be in remarkably close agreement with RMT predictions for the
Gaussian orthogonal ensemble (GOE). Although there have been several other
successful tests of RMT using nuclear resonances, the NDE is perhaps the
most important because, as stated in Ref. \cite{We2009}, \textquotedblleft
As a result of these analyses, it became generally accepted that proton and
neutron resonances in medium weight and heavy nuclei agree with GOE
predictions.\textquotedblright\ Hence the NDE routinely is cited as
providing striking confirmation of RMT. We have carefully analyzed both
neutron-width and neutron resonance-energy data in the NDE and found significant
disagreements between these data and RMT predictions. These analyses, results, 
and an overall assessment of the NDE are described in the next three subsections.

\subsection{Testing the PTD with the NDE}

RMT for the GOE predicts that $\Gamma_{n}^{0}$ data follow the PTD. 
Reduced neutron widths have been reported for a subset of 1245 resonances in
the NDE of Ref. \cite{Bo83}, consisting of 14 to 178 measurements for 24
nuclides. Details of the analyses, nuclides included, primary references, 
number of resonances in each, etc. can be found in Ref. \cite{Ko2011}.

Analyzing a data set comprised of many different nuclides such as the NDE
involves at least two additional potential pitfalls to those mentioned 
in Section \ref{ExpLimits} above. First, apparent
sensitivities of the various experiments from which the NDE was derived
differ by several orders of magnitude. Therefore, if the entire NDE were
analyzed as a single set (as was done in Ref. \cite{Bo83}), to avoid 
systematic error the threshold
must be at least as high as the highest apparent individual threshold.
However, doing this will exclude so many resonances from the analysis that
the statistical precision of the result will be substantially reduced, and
hence at least partially negate the reason for assembling the NDE in the
first place. Therefore, separate thresholds should be used for each of the 
NDE nuclides. Second, the average reduced neutron width 
$\langle \Gamma_{n}^{0}\rangle $ depends on the shape of the distribution, 
and this shape 
may be different for each NDE nuclide. Therefore, it is important to include 
separate $\langle\Gamma _{n}^{0}\rangle$ parameters for each NDE nuclide. 
For these reasons, separate ML analyses must be made for each NDE nuclide, 
and then the combined result subsequently compared to theory.

For the initial analyses, thresholds just below the smallest observed
resonance for each NDE nuclide were used. Because experiment thresholds
might not be precisely sharp, it is expected that the resulting $\nu_{n} $
values would be systematically a bit large. However, almost all $\nu_{n} $
values were less than the PTD value of 1.0. The weighted average of results
at minima thresholds for the 24 nuclides in the NDE is $\nu_{n} =0.801\pm \
0.052 $, which is 3.8 standard deviations smaller than the predicted result
of $\nu_{n} =1$. Hence these data reject the PTD with a statistical significance
of 99.98\%. A $\nu_{n} $ value significantly less than the PTD could be a sign
of interesting physics. However, a more likely explanation is that the NDE
contains sizable \textit{p}-wave contamination.

That the NDE is contaminated by \textit{p}-wave resonances is evident in
many ways. First, many resonances in the NDE have been identified as 
$p$-wave or of uncertain parity. For example, in Ref. \cite{Co78}, 58 
$p$-wave resonances in $^{232}$Th were assigned on the basis of $\gamma $%
-cascade information, 13 of which are in the NDE. One of these 13 resonances
also is known \cite{Fr92,St98} to be $p$-wave by its observed
parity-violating asymmetry. For over half (14/24) of the NDE nuclides we
analyzed, these resonances account for 5\% or more of the total, for 10 of
the 24 they are at least 10\%, and in the three worst cases about 35\%.
Second, as explained in Ref. \cite{Ko2011}, that many of the NDE resonances 
are, in fact, \textit{p} wave is
reinforced by the behavior of the $\nu_{n} $ values from the ML analyses as
functions of threshold. In many cases, $\nu_{n} $ systematically increases with
threshold before gradually stabilizing. This is just the behavior expected
for a population of $s$-wave resonances contaminated by \textit{p}%
-wave resonances.

As explained in Ref. \cite{Ko2011}, 
removing effects of these \textit{p}-wave resonances from the NDE ML
analyses is a simple matter of raising thresholds until they are above the
largest previously identified \textit{p}-wave resonance and/or $\nu_{n} $
stabilizes as a function of threshold. The resulting weighted average for
the NDE is still in conflict with the RMT prediction for the GOE, albeit in
the opposite direction from the result using the lowest thresholds: $\nu_{n}
=1.217\pm 0.092$, corresponding to a confidence level of 98.17\% for
excluding the PTD. Hence, when the NDE is cleansed of \textit{p}-wave
resonances, the data still reject the PTD with high confidence. 
Furthermore, because our $^{147}$Sm and $^{192,194}$Pt results show that
$\nu_{n}$ can be on either side of the PTD value of 1.0, it seems likely that 
any analysis from which an average $\nu_{n}$ is calculated from results for 
several nuclides may underestimate deviations from the PTD.

In the next subsection, we describe our 
analysis of the NDE using the $\Delta _{3}$ statistic. This analysis illustrates 
several more related problems in the NDE.

\subsection{Reexamining $\Delta _{3}$ for the NDE}

The initial claim of excellent agreement between the NDE and RMT was based 
mainly on the $\Delta _{3}$ resonance-spacing statistic. Because our analyses
described above indicated that the neutron-width data in the NDE do not agree with
the PTD, we also examined the $\Delta _{3}$ statistic for this subset of the NDE
data. We will refer to this subset of 1246 neutron resonances as the 
{\em original} NDE, or set ${\cal I}$. There are 1246 energies but only 1245
neutron widths because one resonance in $^{182}$W has a reported energy
but no width.

As mentioned above, several of these NDE resonances have been assigned as 
either definite or probable $p$-wave resonances. Specifically, definite 
$p$-wave assignments, in a primary reference and/or one of the two 
compilations \cite{Mu2006,Su98}, include the following 35 resonance energies 
(in eV): 3333.0 ($^{114}$Cd), 1702.8 
($^{172}$Yb), 1634.6 ($^{174}$Yb), 73135, 100220 and 125130 ($^{64}$Zn), 
196.13, 391.53, 400.86, 411.62, 420.92, 476.30, 540.09, 573.46, 764.7, 820.9, 850.5, 
1114.9, 1204.1, 1372.54, 1387.7, and 1848.6 ($^{232}$Th), and 263.91, 454.1, 555.9, 
732.5, 778.8, 1028.6, 1131.1, 1298.1, 1316.5, 1532.3, 1565.1, 1795.5, and 
2070.9 ($^{238}$U). The original NDE after exclusion of above-specified 35 resonance 
will be referred to as a~{\em corrected} NDE, or subset ${\cal I}_\mathrm{corr}$.

\subsubsection{$\Delta_3$ statistic}

Given a~selected interval of neutron energies $(E_a,E_b)$, the $\Delta_3$ 
statistic is conventionally defined as
\begin{equation}
\Delta_3(L)=\frac{1}{E_b-E_a}\min_{\alpha,\beta}
\int\limits_{E_a}^{E_b}\left[n(E)\!-\!\alpha\!-\!\beta E\right]^2dE.
\label{eq:OriginalDelta3}
\end{equation}
Here, $n(E)$ is equal to the number of the resonances in the interval $(E_a,E_b)$, 
while ${L}=(E_b-E_a)/\mathrm{E}[D]$, where $\mathrm{E}[D]$ is an~expectation 
value of spacing $D$ between energies of neighboring resonances.  

Practical implementation of the statistic according to 
Eq.~(\ref{eq:OriginalDelta3}) faces the problem that, except for the maximum 
value of $L$ for a given set of data, there is no unique definition of $\Delta _{3}$.
That is, except for the maximum value of $L$, there are at least two pairs of 
$E_a$ and $E_b$ for which $\Delta_{3}$ can be calculated. To overcome this difficulty, 
we used a~{\em modified} 
statistic, referred hereafter to as $\Delta^\prime_3$. For $i$-th sequence 
consisting of $\Lambda_i$ energies 
$E_1^{(i)}\leq E_2^{(i)}\leq\cdots\leq  E_{\Lambda_i}^{(i)}$ of consecutive 
resonances of a~given nucleus we introduced a~function 
$\Delta_3^{\prime (i)}(L|\lambda_0)$ of integer argument $L\geq 2$ and integer 
parameter $\lambda_0$ satisfying a~condition that $0<\lambda_0<\Lambda_i-1$. 
Specifically, for $1<L\leq\Lambda_i-\lambda_0+1$
\begin{eqnarray}
\Delta_3^{\prime (i)}(L|\lambda_0)=
\displaystyle\frac{1}{E_{\lambda_0+L-1}^{(i)}-E_{\lambda_0}^{(i)}}
\min_{\alpha,\beta} 
\int\limits_{E_{\lambda_0}^{(i)}}^{E_{\lambda_0+L-1}^{(i)}}
\left[n_i(E)\!-\!\alpha-\beta E\right]^2dE,
\label{eq:Delta3prime}
\end{eqnarray}
while for $L\leq1$ or $L>\Lambda_i-\lambda_0+1$ the function 
$\Delta_3^{\prime (i)}(L|\lambda_0)$ equals zero.

Following the approach of Refs. \cite{Ha82,Bo83}, all the individual 
$\Delta _{3}^{\prime}$ values for a given $L$ were averaged over the 
range of possible $\lambda_0$ values to obtain  
\textit{an~averaged} statistic, $\langle{\Delta^\prime_3}(L)\rangle$, which is 
applicable to any subset ${\cal I}'$ of the full set ${\cal I}$ of the 
available experimental NDE sequences for individual nuclei:
\begin{equation}
\langle{\Delta^\prime_3(L)\rangle=
\frac{\displaystyle\sum\limits_{i\in{\cal I}'}\,\sum\limits_{\lambda_0=1}
^{\Lambda_i-L+1}
\Delta_3^{\prime\,(i)}(L|\lambda_0)}{\displaystyle\sum\limits_{j\in{\cal I}'} 
(\Lambda_j-L+1)}},
\label{eq:AverageDelta3prime}
\end{equation}

Quantities represented by right-hand sides of Eqs. (\ref{eq:OriginalDelta3}) 
and (\ref{eq:Delta3prime}), as well as the 
average quantity defined by Eq. (\ref{eq:AverageDelta3prime}), depend implicitly 
on resonance energies. In the spirit of the 
GOE model, these energies are assumed to be random variables, so that functions 
according to Eqs. (\ref{eq:OriginalDelta3}), (\ref{eq:Delta3prime}) and 
(\ref{eq:AverageDelta3prime}) are indeed statistics. Hereafter, 
depending on the context, we will refer them either to as functions 
or statistics. 

From a~fixed subset 
${\cal I}'\subset{\cal I}$ of the NDE we took the resonance energies and 
determined from them functions $n_i(E)$ for all $i\in{\cal I}'$. Then, 
employing Eqs. (\ref{eq:Delta3prime}) and (\ref{eq:AverageDelta3prime}), 
we determined what we call {\em an~experimental realization} of the averaged 
statistic $\langle\Delta^{\prime\,\mathrm{exp}}_3(L)\rangle$. 

\subsubsection{Calculation of confidence limits}

Although the NDE is a fairly large data set, statistical uncertainties in 
$\langle\Delta^{\prime\,\mathrm{exp}}_3(L)\rangle$ can still be substantial. 
To evaluate these uncertainties and assess whether these data are in agreement 
with theory, we employed a Monte Carlo technique to generate a large number 
(typically 5$\times10^{3}$ to $10^{4}$) of {\em artificial} 
functions $\langle\Delta^\prime_3(L)\rangle$ using energies 
$E_1^{(i)}\leq E_2^{(i)}\leq\cdots\leq  E_{\Lambda_i}^{(i)}$ generated by the 
GOE model for all $i\in{\cal I}' $. In this way, we constructed an 
empirical probability distribution function for $\langle\Delta_3(L)\rangle$ at each 
value of $L$. We compared these simulations to the 
$\langle\Delta^{\prime\,\mathrm{exp}}_3(L)\rangle$ values calculated from the NDE
data to assess whether the NDE is compatible with predictions of the GOE 
model.    

Our implementation of the Monte Carlo technique used 
$N\times N$ GOE matrices. Hence, the $N(N+1)/2$ independent elements of each matrix were  
statistically independent normally distributed random variables with zero 
mean. Variances of the off-diagonal elements were equal to 
a common value $v$, while those of the remaining elements were  equal to $2v$.

Generating energies $E_1^{(i)}\leq E_2^{(i)}\leq\cdots\leq  E_{\Lambda_i}^{(i)}$ 
involved the following steps:

\begin{enumerate}
\item
For a given
$i\in{\cal I}'$ we chose a value $N_{i}>\max(2\,\Lambda_i,100)$, adjusting 
for convenience the variance $v$ to a~value $v_i=1/4N_i$. 
The largest value of $N_i$ used was less than 1000. 
\item
We created a random GOE matrix $\mathrm{A}^{(i)}$ of size $N_i\times N_i$ and 
determined its eigenvalues in ascending order, i.e.
$\epsilon_1^{(i)}\leq \epsilon_2^{(i)}\leq\cdots\leq \epsilon_{N_i}^{(i)}$. 
\item
Using the expression
\begin{equation}
\tilde{\epsilon}^{(i)}_\lambda=
\frac{N_i}{2\pi}\left[2\arcsin\,(\epsilon^{(i)}_\lambda)+
\sin\!\left(2\arcsin(\epsilon^{(i)}_\lambda)\right)\right],
\end{equation}
where $\lambda=1,2,\dots,\Lambda_i$, we transformed the set of eigenvalues 
$\{\epsilon_\lambda^{(i)}\}$ to a modified set 
$\{\tilde{\epsilon}_\lambda^{(i)}\}$. 
This transformation ensures that values of $\{\tilde{\epsilon}_1^{(i)}\}$ 
for $N_i\rightarrow\infty$ are distributed uniformly.
\item
Finally, from a~part of values of $\{\tilde{\epsilon}_1^{(i)}\}$, centered around 
zero, we created neutron energies. For this purpose we adopted a~~prescription 
$E_\lambda^{(i)}=\tilde{\epsilon}_{\lambda_0+\lambda}^{(i)}$ for 
$\lambda=1,2,\dots,\Lambda_i$, where $\lambda_0$ satisfies a~condition that 
$\tilde{\epsilon}^{(i)}_{\lambda_0}\leq-
\Lambda_i/(2N_i)\leq\tilde{\epsilon}^{(i)}_{\lambda_0+1}$ and that the length 
$\Lambda_i$ is equal to that for the $i$-th sequence of the NDE of interest.    
\end{enumerate}  

In this way, we simulated a~large 
number (typically 2$\times10^{3}$ to $10^{4}$) of 
$\langle\Delta^\prime_3(L)\rangle$ values, for which the number and the 
lengths of the sequences are identical to those belonging to either subset 
${\cal I}'\subset {\cal I}$ or ${\cal I}'\subset {\cal I}_\mathrm{corr}$, and from 
the distributions of these values we calculated confidence limits for 
$\langle\Delta^\prime_3(L)\rangle$.

Through further extensive Monte-Carlo simulations we verified that the 
limited sizes of the matrices used resulted in less than 0.2\% systematic 
uncertainty in estimating confidence limits.

\subsubsection{Results} 
\begin{figure}
\begin{minipage}{70mm}
\includegraphics[width=\linewidth]{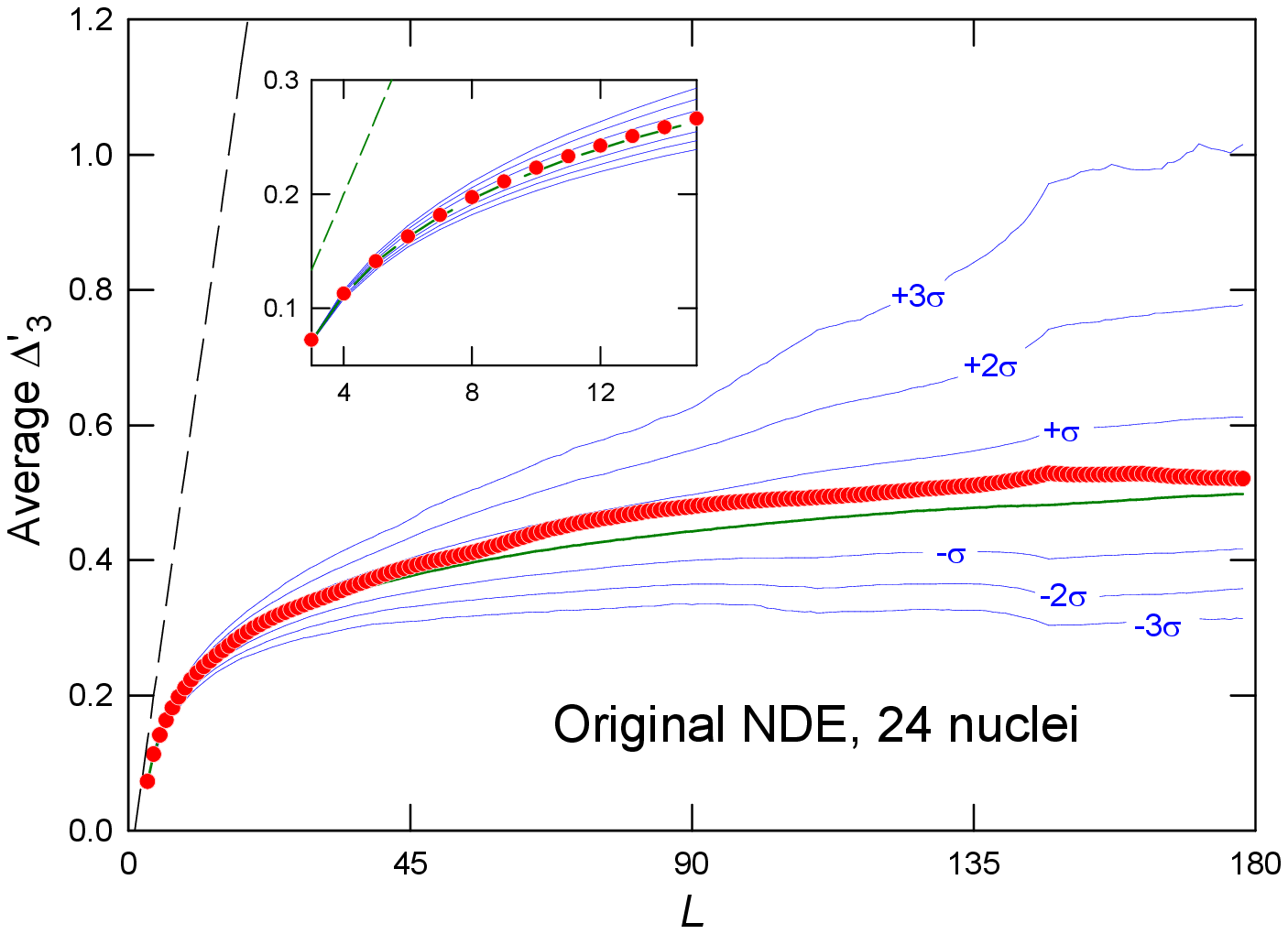}
\caption{Circles: function $\langle\Delta^{\prime\,\mathrm{exp}}_3(L)\rangle$  
for the original NDE. Thick curve: the median of the distribution 
of $\langle\Delta^\prime_3(L)\rangle$ for individual values of $L$ obtained from 
Monte Carlo simulations under the assumption that GOE model holds. Regions 
between pairs of thin curves situated symmetrically to the median curve:
$\pm1\sigma$, $\pm2\sigma$ and $\pm3\sigma$ confidence regions characterizing 
fluctuation of $\langle\Delta^\prime_3(L)\rangle$. Dashed line: function 
$\langle\Delta^\prime_3(L)\rangle$ for the case of uncorrelated neutron 
resonance energies.}
\label{fig:WholeOriginalNDE}
\end{minipage}
\hspace{10mm}
\begin{minipage}{70mm}
\includegraphics[width=\linewidth]{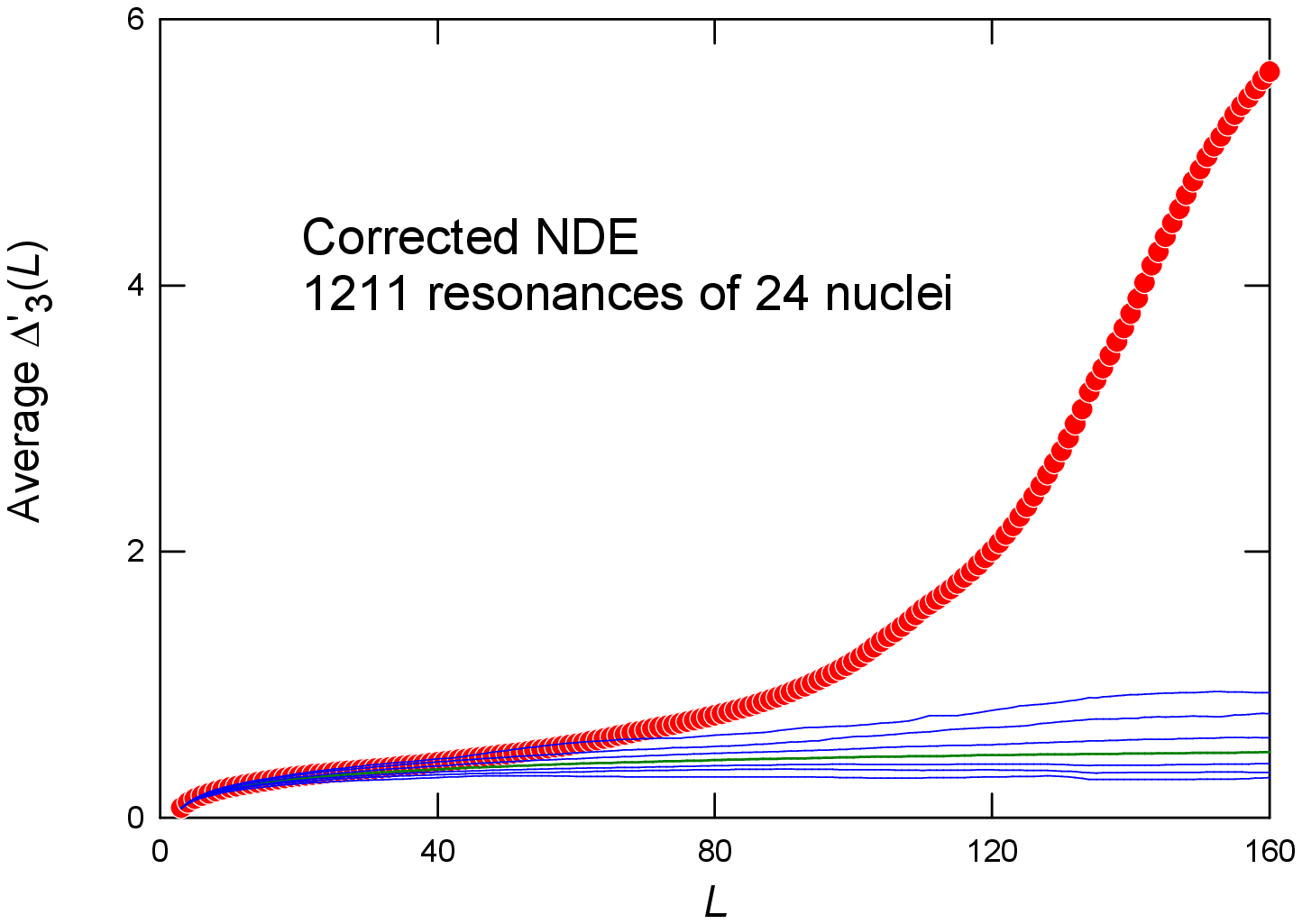}
\caption{Function $\langle\Delta^{\prime\,\mathrm{exp}}_3(L)\rangle$ calculated 
from the corrected NDE. See the caption of 
Fig.~\ref{fig:WholeOriginalNDE} for definitions of the various plots.}
\label{fig:CorrectedNDE}
\end{minipage}
\end{figure}

Results of our analysis for case ${\cal I}'\equiv{\cal I}$ are 
shown in Fig.~\ref{fig:WholeOriginalNDE}, 
where the function $\langle\Delta^{\prime\,\mathrm{exp}}_3(L)\rangle$ is 
plotted together with confidence limits of $\langle\Delta_3(L)\rangle$ . 
This figure indicates that the 
original NDE is in remarkably good agreement with GOE predictions for the 
entire allowed region $L\leq178$, essentially confirming the results claimed
in Ref. \cite{Ha82}.   

\begin{figure}
\begin{minipage}{70mm}
\includegraphics[width=\linewidth]{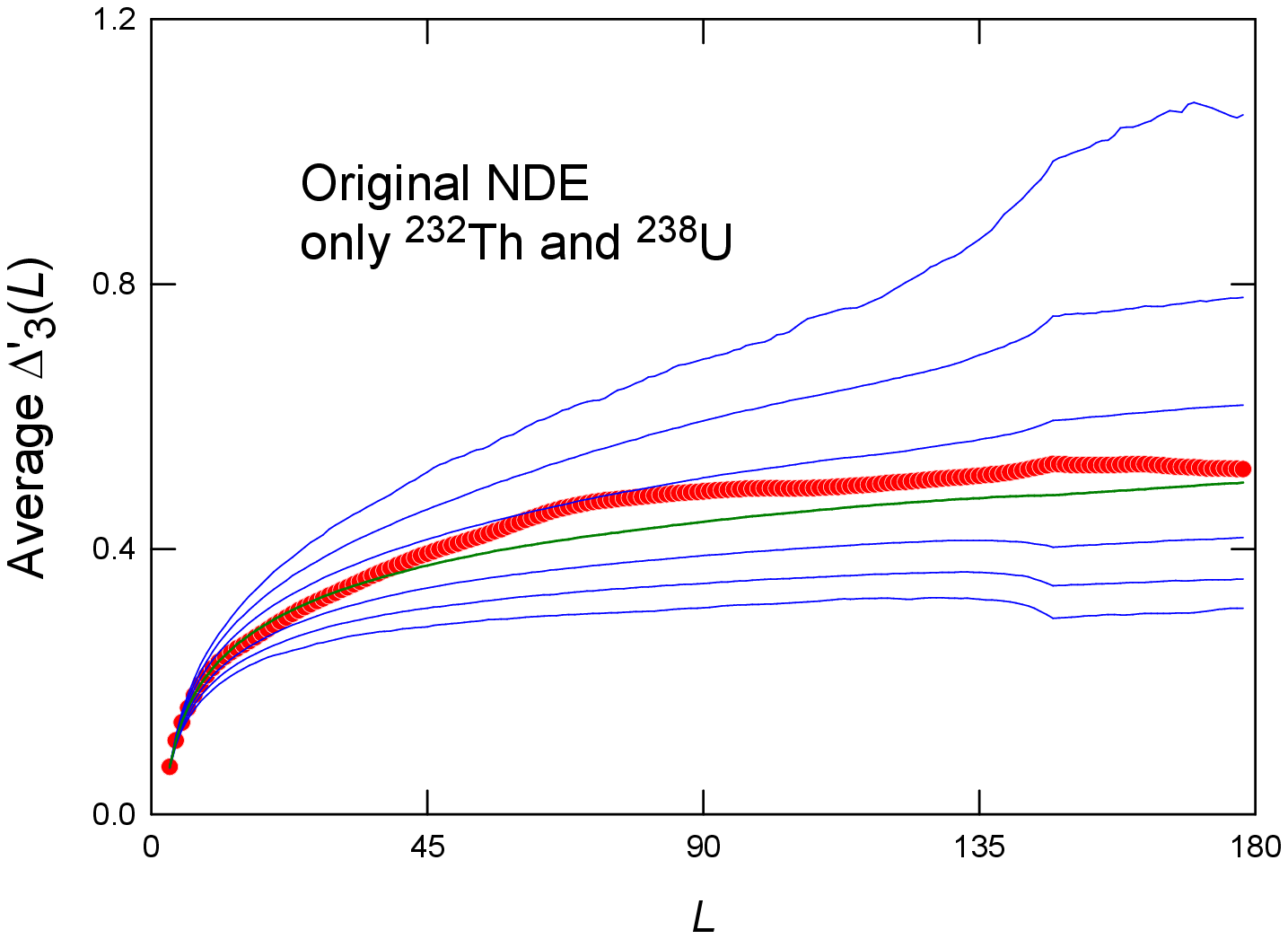}
\caption{Function $\langle\Delta^{\prime\,\mathrm{exp}}_3(L)\rangle$ calculated 
from the original NDE for the subset formed by
$^{232}$Th and $^{238}$U resonances. See the caption of 
Fig.~\ref{fig:WholeOriginalNDE} for definitions of the various plots.}
\label{fig:UThOriginalNDE}
\end{minipage}
\hspace{10mm}
\begin{minipage}{70mm}
\includegraphics[width=\linewidth]{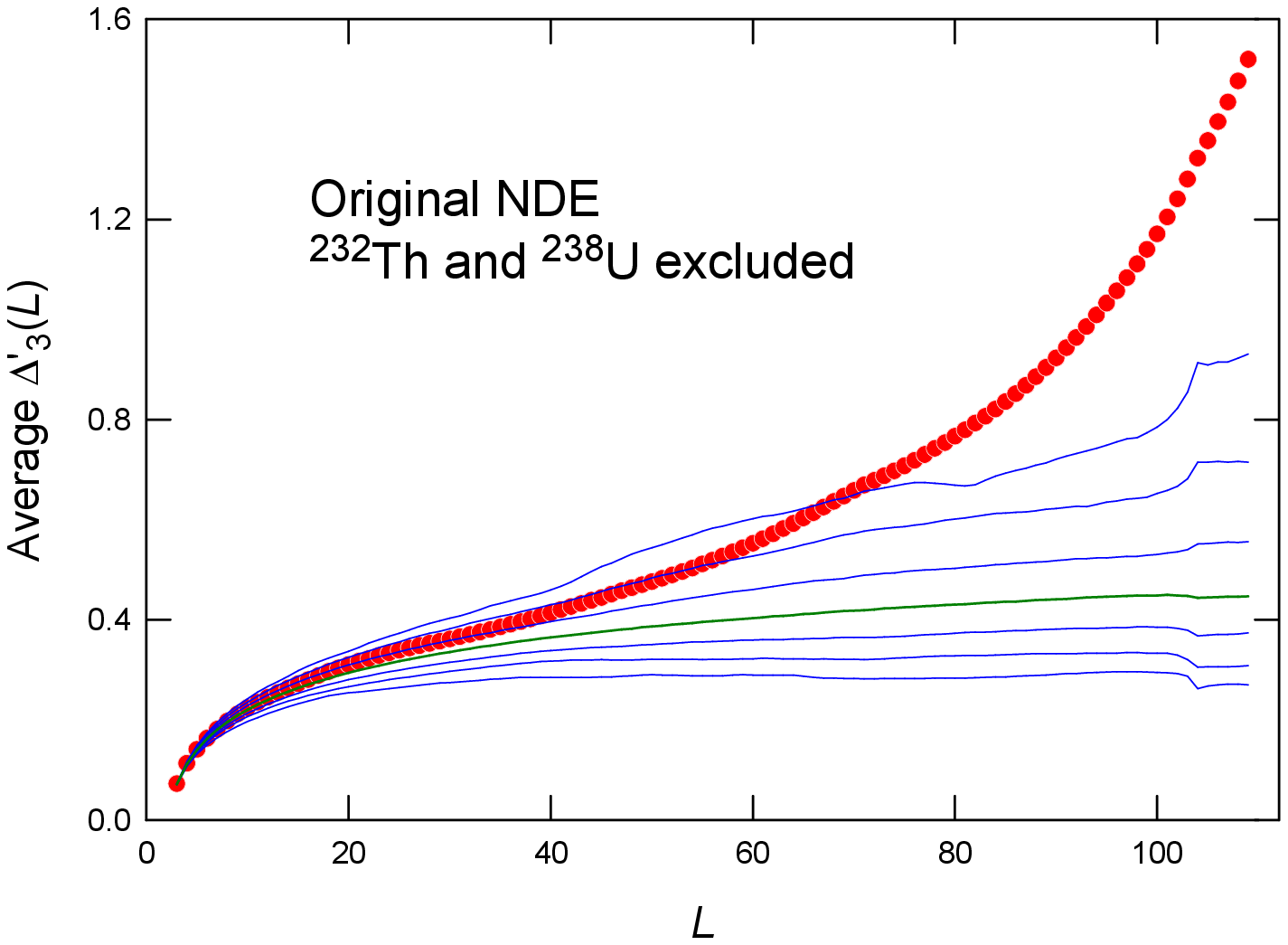}
\caption{Function $\langle\Delta^{\prime\,\mathrm{exp}}_3(L)\rangle$ deduced 
from the original NDE after eliminating the data belonging to $^{232}$Th and 
$^{238}$U. See the caption of 
Fig.~\ref{fig:WholeOriginalNDE} for definitions of the various plots.}
\label{fig:UThExludedOriginalNDE}
\end{minipage}
\end{figure}

The longest sequences of neutron resonance energies in the original NDE 
belong to $^{232}$Th and $^{238}$U. Results of our analysis limited 
to these two nuclides are illustrated in Fig.~\ref{fig:UThOriginalNDE}. Again, 
agreement between the data and GOE predictions is excellent.

However, when $^{232}$Th and $^{238}$U 
are removed from the original NDE, the agreement disappears, as shown in 
Fig.~ \ref{fig:UThExludedOriginalNDE}. In this case, validity of the GOE 
model can be rejected  
with a confidence level greater than 99.99\%. This 
conclusion is not at variance with the results shown in 
Figs.~\ref{fig:WholeOriginalNDE} and \ref{fig:UThOriginalNDE}, but reflects the 
large weights with which $^{232}$Th and $^{238}$U, by virtue of their comparatively 
long sequences, contribute to $\langle\Delta^{\prime\,\mathrm{exp}}_3(L)\rangle$. 
For example, for $L>70$ this weight is greater than 80\%. 

On the other hand, regarding 
the average statistic $\langle\Delta^\prime_3(L)\rangle$, the data from the 
original NDE {\em as a~whole} are inconsistent: one subset of 
these data is in good 
agreement with the GOE model (see Fig.~\ref{fig:UThOriginalNDE}), 
while the complement to this subset is in sharp disagreement 
(Fig.~\ref{fig:UThExludedOriginalNDE}). This inconsistency was 
not be revealed before, as previous analyses \cite{Ha82,Bo83} were not undertaken 
separately for subsets of the NDE.  

To ascertain whether the results shown in Fig. \ref{fig:UThExludedOriginalNDE} 
might be due to the inclusion of $p$-wave resonances, we analyzed the data from 
the {\em corrected} NDE after 
additional exclusion of $^{232}$Th and $^{238}$U. 
Results of the analysis led to 
the same conclusion as in the previous case: the validity of the 
GOE model could be rejected with statistical significance of $>$99.99\%. 
As the excluded $p$-wave resonances accounted for  
only 1.5\% of the total number of resonances (of the {\em corrected} NDE, excluding 
$^{232}$Th and $^{238}$U), this result was not too surprising.

Results of analyzing the full 
{\em corrected} NDE, are illustrated in Fig.~\ref{fig:CorrectedNDE}. 
As can be seen from this figure, exclusion of firm and probable $p$-wave 
resonances from the NDE leads to categoric disagreement with 
predictions of the GOE model for $L$ values as small as 70. This 
disagreement persists even if analysis is further restricted to only  
$^{232}$Th and $^{238}$U resonances of the corrected NDE, with 
unacceptable disagreement again beginning at $L\approx70$.    

It could be argued that the disagreement shown in 
Fig.~\ref{fig:CorrectedNDE} is due to missing $s$-wave resonances 
and/or inclusion of $p$-wave resonances in the corrected 
NDE. It is well known that it becomes increasingly difficult to 
correctly determine resonance parity with increasing energy. 
Therefore, we repeated the analysis of the corrected NDE after shortening each sequence 
by factor of 0.5. In this case we found that at $L\approx72$
the average quantity $\langle\Delta_3^{\prime\,\mathrm{exp}}(L)\rangle$ still 
deviates appreciably 
from its expectation value, by $\approx3.2\sigma$. Additional Monte Carlo 
simulations led to the conclusion that within the GOE model the probability  
$\langle\Delta_3(L)\rangle$ exceeds a confidence limit of 
3.2$\sigma$ at one value of $L$ is still very small, 0.04~\%. So, even 
with such a drastic reduction the data do not support the GOE 
model. Acceptable agreement was achieved only at values $L<60$. 

\subsection{Overall assessment of the NDE}

Given that our analyses of neutron widths as well as energies for the 
same NDE resonances reveal many disagreements with predictions of RMT for 
the GOE, it is obvious to ask why the full NDE was found to
be \cite{Ha82,Bo83} in excellent agreement with this theory. As detailed in
Ref. \cite{Ko2011}, the most likely explanation has to do with the fact that
many of the data in the NDE were selected using this same theory. Almost all
the neutron data in the NDE were obtained by the group at Columbia
University. In their papers, they readily admit that they did not have any
specific tests for separating $s$- from $p$-wave resonances.
Hence, they routinely used measures derived from RMT for the GOE to perform
these separations. Given the numbers of total and assigned $s$-wave
resonances involved, there typically were astronomically large 
(e.g. $10^{27}$ in the case of $^{232}$Th \cite{Ko2011}) numbers of
possible \textquotedblleft $s$-wave\textquotedblright\ sets from
which to choose. Hence, it does not seem surprising that at least one set
could be found which agrees with the statistics predicted by RMT for the GOE
tested in Refs. \cite{Ha82,Bo83}. In other words, the NDE likely demonstrates that it 
is possible to take an incomplete data set and make it agree with GOE statistics 
by also making it impure.

With the experiment techniques available at the time, 
it was just not possible to obtain a complete and pure set of $s$-wave 
resonances, especially considering that many of the NDE nuclides are very 
near the peaks of the $p$- and valleys of the $s$-wave neutron strength 
functions. Even with the improvements in experiments over the intervening 40 
years, it still is not possible to obtain such a pure and complete set of 
$s$-wave neutron resonances as the NDE is purported to be. The main problem 
remains the reliable separation of small (neutron width) $s$- from large $p$-wave 
resonances. In section \ref{NewExpTechs}, we briefly describe some new 
experiment techniques that have greatly improved the ability to make firm 
resonance spin and parity assignments. Application of these techniques to 
nuclides near peaks of the $s$-wave neutron strength function should provide 
much better data.

\section{Other reported deviations from the PTD}\label{OtherExps}

As noted above, data for the NDE, $^{147}$Sm, and $^{192,194}$Pt all exclude
the PTD with high confidence. In addition, because the $^{147}$Sm and $%
^{192,194}$Pt data demonstrate that $\nu_{n} $ can be on either side of the PTD
value of $\nu_{n} =1$, the NDE result, which combines analyses of 24 different
nuclides, likely underestimates the probability with which the PTD can be
excluded. There have been a few other reported significant deviations from
the PTD of which we are aware.

There have been several reports \cite{Ri69,Fo71,Fo71a,Ra72} that the lowest
energy neutron resonances in $^{232}$Th deviate strongly from the PTD. Our
ML analysis \cite{Ko2009a} of the data verifies these reports; the data
change from disagreeing ($\nu_{n} =3.8\pm 1.3$) to agreeing ($\nu_{n} =0.83\pm 0.68$%
) with the PTD over the first two groups of 25 resonances. Hence, data
for both $^{232}$Th and $^{147}$Sm indicate that the shape of the $\Gamma
_{n}^{0}$ distribution can change in a relatively narrow energy region. 
 As described in Ref. \cite{Ko2012b}, our Pt 
data also suggest that $\nu_{n}$ may depend on energy.

The combined data for five odd-A nuclides also were found \cite{Ca76} to
deviate strongly from the PTD despite the fact that very few resonances
apparently had been missed. As was the case for $^{147}$Sm and $^{232}$Th, $%
\nu_{n} $ was found to be significantly greater than 1.0 for these nuclides.

\section{Brief overview of theory}\label{Theory}

Our Pt results have inspired a number of theoretical papers \cite%
{We2010,Ce2011,Vo2011}, some of which have their roots in earlier
publications (e.g. see Ref. \cite{Kl85}). Model calculations \cite{Al92}
also have shown that deviations from the PTD may sometimes occur even when
spacing statistics such as $\Delta _{3}$ do not deviate from GOE
predictions. However, these latter calculations were for levels near the
ground state where collective effects can be important and hence leave
unexplained how such effects can manifest themselves at the relatively high
excitations corresponding to typical neutron separation energies.

In Ref. \cite{We2010} it was proposed that the standard transformation from
measured to reduced neutron width, 
$\Gamma _{n}^{0}$=$\Gamma _{n}/\sqrt{E_n/\mathrm{1 eV}}$, may be different 
for nuclides near the peaks of the $s$-wave
neutron strength function. However, analysis \cite{Ko2010b} of our Pt data
showed that the new transformation resulted in increased deviations from the
PTD. In addition, this proposal could not explain cases such as $^{232}$Th
which are near minima in the $s$-wave neutron strength function.

Strictly speaking, RMT predictions are valid only for bound states. Possible
deviations related to the \textquotedblleft openness\textquotedblright\ of
the system above the particle separation energy were, to our knowledge,
first explored in Ref. \cite{Kl85}. Recent models in this vein are described
in Refs. \cite{Ce2011,Vo2011}. The model of Ref. \cite{Ce2011} does not
appear to be able to explain our Pt result because the required coupling
parameter is $\sim 1000$ times larger than that estimated from our data. The
model of Ref. \cite{Vo2011} results in width distributions broader than the
PTD, in agreement with our Pt data. However, it is unclear whether
significant deviations from the PTD would remain if more realistic matrix
elements were used. This model also predicts fluctuations
larger than the PTD for electromagnetic transitions, in qualitative
agreement with our $\Gamma_{\gamma}$ data discussed above.

As far as we know, no published model can, in general, explain $\nu_{n}$
values greater than 1.0 (although the model of Ref. \cite{We2010} can result
in $\nu _{n}>1$ under certain conditions) or large changes in $\nu _{n}$ or $%
\nu _{\gamma }$ within relatively narrow energy regions as observed in $%
^{147}$Sm and $^{232}$Th, and suggested \cite{Ko2012b} by our Pt data.

\section{\label{NewExpTechs} New experimental techniques}

New techniques \cite{Ko2011a,Ko2007,Be2011,Ba2012} for determining
spins and parities of neutron resonances should improve the separation of
resonances with differing spins and parities and hence help to overcome what
has been the largest barrier to better tests of the PTD. All these new 
methods make use of information contained in the $\gamma$ cascade 
following neutron capture. It is an old idea \cite{Co68}, but 
improvements in neutron sources, detectors, and data acquisition 
apparatus has led to large improvements over previous results. For 
example, in the case of $^{95}$Mo neutron resonances, applying 
these methods with a single pair of low-efficiency C$_{6}$D$_{6}$ 
detectors increased the number of firm resonance $J^{\pi}$ 
assignments from 32 \cite{Sh2007} to 218 \cite{Ko2011a} in this very difficult 
case.

\section{Summary and Conclusions}\label{Conclusions}

We have found significant disagreements with predictions of RMT for the GOE 
in neutron-resonance data for $^{147}$Sm, $^{192,194}$Pt, $^{232}$Th, and 
the NDE. The main barrier to more and better tests of theory is reliable 
separation of $s$- from $p$-wave resonances, but new techniques 
appear to be well on their way to surmounting this barrier. For neutron 
resonances, tests involving the PTD appear to be the most sensitive and 
reliable, largely because experiment limitations can be incorporated into 
the statistical methods in a straightforward manner. Tests involving 
resonance-spacing statistics such as those in Refs. \cite{Ha82,Bo83} 
appear to have ignored these limitations, and the good agreement between 
data and theory claimed appears to be due to the fact that many of the 
data used were selected using the theory being tested. Other tests that are 
purported to be less sensitive to missing or spurious resonances have been 
proposed \cite{Le86}, but have been shown \cite{By96} to lead to ambiguous 
results for sample sizes typical of nuclear data.

This work was supported by the Office of Nuclear Physics of the U.S. Department
of Energy under Contract No. DE-AC05-00OR22725 with UT-Battelle, LLC, and by 
Czech Research Plans MSM-021620859 and INGO-LA08015. This
work has benefited from the use of the LANSCE facility at Los Alamos
National Laboratory which was funded by the U.S. Department of Energy and
currently is operated by Los Alamos National Security, LLC, under contract 
DE-AC52-06NA25396.

\bibliographystyle{fdp}
\bibliography{ACOMPAT,pauls}

\begin{thebibliography}{[10]}

\bibitem{Po56}
 \textsc{C.\,E. Porter} and  \textsc{R.\,G. Thomas},
 \jr{Phys. Rev.} \textbf{104}, 483 (1956).


\bibitem{We2009}
 \textsc{H.\,A. Weidenm{\"u}ller} and  \textsc{G.\,E. Mitchell},
 \jr{Rev. Mod. Phys.} \textbf{81}, 539 (2009).


\bibitem{Ko2007}
 \textsc{P.\,E. Koehler},  \textsc{J.\,L. Ullmann},  \textsc{T.\,A. Bredeweg},
  \textsc{J.\,M. O'Donnell},  \textsc{R.~Reifarth},  \textsc{R.\,S. Rundberg},
  \textsc{D.\,J. Vieira},  and  \textsc{J.\,M. Wouters},
 \jr{Phys. Rev. C} \textbf{76}, 025804 (2007).


\bibitem{Ko2010a}
 \textsc{P.\,E. Koehler},  \textsc{F.~Be\v{c}v{\'a}\v{r}},
  \textsc{M.~Krti\v{c}ka},  \textsc{J.\,A. Harvey},  and  \textsc{K.\,H.
  Guber},
 \jr{Phys. Rev. Lett.} \textbf{105}, 072502 (2010).


\othercit
\bibitem{Pe82}
 \textsc{R.\,W. Peelle},  \textsc{J.\,A. Harvey},  \textsc{F.\,C. Maienschein},
   \textsc{L.\,W. Weston},  \textsc{D.\,K. Olsen},  \textsc{D.\,C. Larson},
  and  \textsc{R.\,L. Macklin},
Neutron research and facility development at the {O}ak {R}idge electron linear
  accelerator 1970-1995,
Tech. Rep. ORNL/TM-8225, Oak Ridge National Laboratory, 1982.


\bibitem{Li90}
 \textsc{P.\,W. Lisowski},  \textsc{C.\,D. Bowman},  \textsc{G.\,J. Russell},
  and  \textsc{S.\,A. Wender},
 \jr{Nucl. Sci. Eng.} \textbf{106}, 208 (1990).


\bibitem{Ko2012}
 \textsc{P.\,E. Koehler},  \textsc{R.~Reifarth},  \textsc{J.\,L. Ullmann},
  \textsc{T.\,A. Bredeweg},  \textsc{J.\,M. O'Donnell},  \textsc{R.\,S.
  Rundberg},  \textsc{D.\,J. Vieira},  and  \textsc{J.\,M. Wouters},
 \jr{Phys. Rev. Lett.} \textbf{??}, ?? (2012).


\bibitem{Ko2011}
 \textsc{P.\,E. Koehler},
 \jr{Phys. Rev. C} \textbf{84}, 034312 (2011).


\bibitem{Ha82}
 \textsc{R.\,U. Haq},  \textsc{A.~Pandey},  and  \textsc{O.~Bohigas},
 \jr{Phys. Rev. Lett.} \textbf{48}, 1086 (1982).


\othercit
\bibitem{Bo83}
 \textsc{O.~Bohigas},  \textsc{R.\,U. Haq},  and  \textsc{A.~Pandey},
Fluctuation properties of nuclear energy levels and widths: Comparison of
  theory with experiment,
 in: Nuclear Data for Science and Techology, edited by K.\,H. Bockhoff (D.
  Reidel, Dordrecht, 1983),  p.\,809.


\othercit
\bibitem{Ri69}
 \textsc{P.~Ribon},
Etudes de Quelques Proprietes Des Niveaux Excites Des Noyaux Composes Formes
  Par L'interaction de Neutrons Lents Avec $^{103}$Rh, Xe, $^{155,157}$Gd, and
  $^{232}$Th,
PhD thesis, Universite de Paris, 1969.


\othercit
\bibitem{Fo71}
 \textsc{L.~Forman},  \textsc{A.\,D. Schelberg},  \textsc{J.\,H. Warren},
  \textsc{M.\,V. Harlow},  \textsc{H.\,A. Grench},  and  \textsc{N.\,W. Glass},
Thorium-232 neutron capture in the region 20 e{V} - 30 ke{V},
 in: Proceedings of the Third Conference on Neutron Cross Sections and
  Technology, edited by J.\,A. Harvey and R.\,L. Macklin (National Technical
  Information Service, U. S. Dept. of Commerce, Springfield, Virginia, 1971),
  p.\,735.


\bibitem{Fo71a}
 \textsc{L.~Forman},  \textsc{A.\,D. Schelberg},  \textsc{J.\,H. Warren},  and
  \textsc{N.\,W. Glass},
 \jr{Phys. Rev. Lett.} \textbf{27}, 117 (1971).


\bibitem{Ra72}
 \textsc{F.~Rahn},  \textsc{H.\,S. Camarda},  \textsc{G.~Hacken},
  \textsc{J.~W.~W.~Havens},  \textsc{H.\,I. Liou},  \textsc{J.~Rainwater},
  \textsc{M.~Slagowitz},  and  \textsc{S.~Wynchank},
 \jr{Phys. Rev. C} \textbf{6}, 1854 (1972).


\bibitem{Ca76}
 \textsc{R.\,F. Carlton},  \textsc{S.~Raman},  \textsc{J.\,A. Harvey},  and
  \textsc{G.\,G. Slaughter},
 \jr{Phys. Rev. C} \textbf{14}, 1439 (1976).


\bibitem{We2010}
 \textsc{H.\,A. Weidenm{\"u}ller},
 \jr{Phys. Rev. Lett.} \textbf{105}, 232501 (2010).


\bibitem{Ce2011}
 \textsc{G.\,L. Celardo},  \textsc{N.~Auerbach},  \textsc{F.\,M. Izrailev},
  and  \textsc{V.\,G. Zelevinsky},
 \jr{Phys. Rev. Lett.} \textbf{106}, 042501 (2011).


\bibitem{Vo2011}
 \textsc{A.~Volya},
 \jr{Phys. Rev. C} \textbf{83}, 044312 (2011).


\othercit
\bibitem{Sh2012}
 \textsc{G.~Shchedrin} and  \textsc{V.~Zelevinsky},
Resonance width distribution for open quantum systems, 2012,
arXiv:1112.4919v2.


\bibitem{Kl85}
 \textsc{P.~Kleinw{\"a}chter} and  \textsc{I.~Rotter},
 \jr{Phys. Rev. C} \textbf{32}, 1742 (1985).


\bibitem{Re2004}
 \textsc{R.~Reifarth},  \textsc{T.\,A. Bredeweg},  \textsc{A.~Alpizar-Vicente},
   \textsc{J.\,C. Browne},  \textsc{E.\,I. Esch},  \textsc{U.~Greife},
  \textsc{R.\,C. Haight},  \textsc{R.~Hatarik},  \textsc{A.~Kronenberg},
  \textsc{J.\,M. O'Donnell},  \textsc{R.\,S. Rundberg},  \textsc{J.\,L.
  Ullmann},  \textsc{D.\,J. Vieira},  \textsc{J.\,B. Wilhelmy},  and
  \textsc{J.\,M. Wouters},
 \jr{Nucl. Instrum. Methods in Phys. Res. A} \textbf{531}, 530 (2004).


\bibitem{Be2011}
 \textsc{F.~Be\v{c}v{\'a}\v{r}},  \textsc{P.\,E. Koehler},
  \textsc{M.~Krti\v{c}ka},  \textsc{G.\,E. Mitchell},  and  \textsc{J.\,L.
  Ullmann},
 \jr{Nucl. Instrum. Methods Phs. Res. A} \textbf{647}, 73 (2011).


\bibitem{Ko2004}
 \textsc{P.\,E. Koehler},  \textsc{Y.\,M. Gledenov},  \textsc{T.~Rauscher},
  and  \textsc{C.~Fr\={o}hlich},
 \jr{Phys. Rev. C} \textbf{69}, 015803 (2004).


\bibitem{Be98}
 \textsc{F.~Be\v{c}v{\'a}\v{r}},
 \jr{Nucl. Instr. Methods Phys. Res. A} \textbf{417}, 434 (1998).


\bibitem{Ko2011a}
 \textsc{P.\,E. Koehler},  \textsc{F.~Becvar},  \textsc{J.\,A. Harvey},
  \textsc{M.~Krticka},  and  \textsc{K.\,H. Guber},
 \jr{J. Korean Phys. Soc.} \textbf{59}, 2088 (2011).


\othercit
\bibitem{Co80}
 \textsc{W.\,J. Conover},
Practical Nonparametric Statistics (John Wiley and Sons, New York, 1980).


\bibitem{As66}
 \textsc{M.~Asghar},  \textsc{C.\,M. Chaffey},  \textsc{M.\,C. Moxon},
  \textsc{N.\,J. Pattenden},  \textsc{E.\,R. Rae},  and  \textsc{C.\,A.
  Uttley},
 \jr{Nucl. Phys.} \textbf{76}, 196 (1966).


\othercit
\bibitem{Co78}
 \textsc{F.~Corvi},  \textsc{G.~Pasquariello},  and  \textsc{T.\,V. der Veen},
$p$-wave assignment of $^{232}${T}h neutron resonances,
 in: Neutron Physics and Nuclear Data for Reactors and Other Applied Purposes,
  edited by C.~Kousnetzoff (Organisation for Economic Co-operation and
  Development, Paris, 1978),  p.\,712.


\bibitem{Fr92}
 \textsc{C.\,M. Frankle},  \textsc{J.\,D. Bowman},  \textsc{J.\,E. Bush},
  \textsc{P.\,P.\,J. Delheij},  \textsc{C.\,R. Gould},  \textsc{D.\,G. Haase},
  \textsc{J.\,N. Knudson},  \textsc{G.\,E. Mitchell},  \textsc{S.~Penttila},
  \textsc{H.~Postma},  \textsc{N.\,R. Roberson},  \textsc{S.\,J. Seestrom},
  \textsc{J.\,J. Szymanski},  \textsc{S.\,H. Yoo},  \textsc{V.\,W. Yuan},  and
  \textsc{X.~Zhu},
 \jr{Phys. Rev. C} \textbf{46}, 778 (1992).


\bibitem{St98}
 \textsc{S.\,L. Stephenson},  \textsc{J.\,D. Bowman},  \textsc{B.\,E.
  Crawford},  \textsc{P.\,P.\,J. Delheij},  \textsc{C.\,M. Frankle},
  \textsc{M.~Iinuma},  \textsc{J.\,N. Knudson},  \textsc{L.\,Y. Lowie},
  \textsc{A.~Masaike},  \textsc{Y.~Matsuda},  \textsc{G.\,E. Mitchell},
  \textsc{S.\,I. Penttila},  \textsc{H.~Postma},  \textsc{N.\,R. Roberson},
  \textsc{S.\,J. Seestrom},  \textsc{E.\,I. Sharapov},  \textsc{Y.\,F. Yen},
  and  \textsc{V.\,W. Yuan},
 \jr{Phys. Rev. C} \textbf{58}, 1236 (1998).


\othercit
\bibitem{Mu2006}
 \textsc{S.\,F. Mughabghab},
Atlas of Neutron Resonances: Resonance Parameters and Thermal Cross Sections
  Z=1-100 (Elsevier, Amsterdam, The Netherlands, 2006).


\othercit
\bibitem{Su98}
 \textsc{S.\,I. Sukhoruchkin},  \textsc{Z.\,N. Soroko},  and  \textsc{V.\,V.
  Deriglazov},
Low Energy Neutron Physics (Springer-Verlag, Berlin, 1998).


\othercit
\bibitem{Ko2009a}
 \textsc{P.\,E. Koehler},
Reduced neutron widths in the nuclear data ensemble: Experiment and theory do
  not agree,
 in: {CNR*09} Second International Workshop on Compound Nuclear Reactions and
  Related Topics, edited by L.~Bonneau, N.~Dubray, F.~Gunsing,  and B.~Jurado
  (EPJ Web of Conferences, Paris, 2009),  p.\,05001.


\othercit
\bibitem{Ko2012b}
 \textsc{P.\,E. Koehler},
$\nu$ we never knew you,
 in: 14$^{th}$ International Symposium on Capture Gamma-Ray Spectroscopy and
  Related Topics, edited by P.~Garrett (World Scientific, Singapore, 2012),
  p.\,??


\bibitem{Al92}
 \textsc{Y.~Alhassid} and  \textsc{A.~Novoselsky},
 \jr{Phys. Rev. C} \textbf{45}, 1677 (1992).


\othercit
\bibitem{Ko2010b}
 \textsc{P.\,E. Koehler},  \textsc{F.~Becvar},  \textsc{M.~Krticka},
  \textsc{H.\,A. Harvey},  and  \textsc{K.\,H. Guber},
Comment on "{D}istribution of partial neutron widths for nucldi close to
  maximum of the neutron strength function, 2010,
http://arxiv.org/abs/1101.4533.


\bibitem{Ba2012}
 \textsc{B.~Baramsai},  \textsc{G.\,E. Mitchell},  \textsc{U.~Agvaanluvsan},
  \textsc{F.~Becvar},  \textsc{T.\,A. Bredeweg},  \textsc{A.~Chyzh},
  \textsc{A.~Couture},  \textsc{D.~Dashdorj},  \textsc{R.\,C. Haight},
  \textsc{M.~Jandel},  \textsc{A.\,L. Keksis},  \textsc{M.~Krticka},
  \textsc{J.\,M. O'Donnell},  \textsc{R.\,S. Rundberg},  \textsc{J.\,L.
  Ullmann},  \textsc{D.\,J. Vieira},  and  \textsc{C.\,L. Walker},
 \jr{Phys. Rev. C} \textbf{85}, 024622 (2012).


\bibitem{Co68}
 \textsc{C.~Coceva},  \textsc{F.~Corvi},  \textsc{P.~Giacobbe},  and
  \textsc{C.~Carraro},
 \jr{Nucl. Phys.} \textbf{A117}, 586 (1968).


\bibitem{Sh2007}
 \textsc{S.\,A. Sheets},  \textsc{U.~Agvaanluvsan},  \textsc{J.\,A. Becker},
  \textsc{F.~Be\v{c}v{\'a}\v{r}},  \textsc{T.\,A. Bredeweg},  \textsc{R.\,C.
  Haight},  \textsc{M.~Krti\v{c}ka},  \textsc{M.~Jandel},  \textsc{G.\,E.
  Mitchell},  \textsc{J.\,M. O'Donnell},  \textsc{W.\,E. Parker},
  \textsc{R.~Reifarth},  \textsc{R.\,S. Rundberg},  \textsc{E.\,I. Sharapov},
  \textsc{I.~Tomandl},  \textsc{J.\,L. Ullmann},  \textsc{D.\,J. Vieira},
  \textsc{J.\,M. Wouters},  \textsc{J.\,B. Wilhelmy},  and  \textsc{C.\,Y.
  Wu},
 \jr{Phys. Rev. C} \textbf{76}, 064317 (2007).


\bibitem{Le86}
 \textsc{L.~Leviandier},  \textsc{M.~Lombardi},  \textsc{R.~Jost},  and
  \textsc{J.\,P. Pique},
 \jr{Phys. Rev. Lett.} \textbf{56}, 2449 (1986).


\bibitem{By96}
 \textsc{C.\,R. Bybee},  \textsc{G.\,E. Mitchell},  and
  \textsc{J.~J.~F.~Shriner},
 \jr{Z. Phys. A} \textbf{355}, 327 (1996).


\end{thebibliography}

\newif\ifabfull\abfulltrue
\providecommand{\WileyBibTextsc}{}
\let\textsc\WileyBibTextsc
\providecommand{\othercit}{}
\providecommand{\jr}[1]{#1}
\providecommand{\etal}{~et~al.}

\end{document}